\documentclass[sigconf]{acmart}

\AtBeginDocument{%
  \providecommand\BibTeX{{%
    \normalfont B\kern-0.5em{\scshape i\kern-0.25em b}\kern-0.8em\TeX}}}

\usepackage{multirow}
\usepackage{hyperref}

\copyrightyear{2021}
\acmYear{2021}
\setcopyright{none}
\acmConference[LAK21]{LAK21: 11th International Learning Analytics and Knowledge Conference}{April 12--16, 2021}{Irvine, CA, USA}
\acmBooktitle{LAK21: 11th International Learning Analytics and Knowledge Conference (LAK21), April 12--16, 2021, Irvine, CA, USA}
\acmDOI{10.1145/3448139.3448173}
\begin{document}
\title{Learning Skill Equivalencies Across Platform Taxonomies}

\author{Zhi Li}
\affiliation{%
  \institution{University of California, Berkeley}
}
\email{zhili@berkeley.edu}

\author{Cheng Ren}
\affiliation{%
  \institution{University of California, Berkeley}
}
\email{cheng.ren@berkeley.edu}

\author{Xianyou Li}
\affiliation{%
  \institution{University of California, Berkeley}
}
\email{lixianyou@berkeley.edu}

\author{Zachary A. Pardos}
\affiliation{%
  \institution{University of California, Berkeley}
}
\email{pardos@berkeley.edu}


\begin{abstract}
Assessment and reporting of skills is a central feature of many digital learning platforms. With students often using multiple platforms, cross-platform assessment has emerged as a new challenge. While technologies such as Learning Tools Interoperability (LTI) have enabled communication between platforms, reconciling the different skill taxonomies they employ has not been solved at scale. In this paper, we introduce and evaluate a methodology for finding and linking equivalent skills between platforms by utilizing problem content as well as the platform's clickstream data. We propose six models to represent skills as continuous real-valued vectors, and leverage machine translation to map between skill spaces. The methods are tested on three digital learning platforms: ASSISTments, Khan Academy, and Cognitive Tutor. Our results demonstrate reasonable accuracy in skill equivalency prediction from a fine-grained taxonomy to a coarse-grained one, achieving an average recall@5 of 0.8 between the three platforms. Our skill translation approach has implications for aiding in the tedious, manual process of taxonomy to taxonomy mapping work, also called crosswalks, within the tutoring as well as standardized testing worlds.
\end{abstract}


\keywords{Skill equivalencies, transfer models, crosswalks, taxonomies, digital learning platforms, representation learning, machine translation, interoperability, acknowledging prior knowledge, app hand-offs.}

\maketitle

\section{Introduction}

Digital learning platforms commonly tag skills to assessment items in order to measure students' mastery and guide their learning trajectories \cite{koedinger2012knowledge,bier2014approach,razzaq2007developing}. Due to the formative nature of digital learning platforms \cite{heffernan2014assistments}, their skill taxonomies are generally finer-grained as compared to those found in large scale summative tests, where broader constructs and abilities are measured \cite{wilson2004constructing}. The presence of an accurate skill or knowledge component model \cite{koedinger2012knowledge} in a digital learning platform can have a significant positive impact on its efficacy \cite{cen2007over}. Conversely, the absence of an accurate expert skill model can impede effective assessment of mastery and subsequently prohibit adaptive learning approaches \cite{pardos2013adapting}.

Customarily, digital learning platforms have developed their own taxonomies, but the demand for greater interoperability across platforms is rapidly growing \cite{porcello2013crowdsourcing}. Teachers and students today often use a mix of devices and many learning platforms per device in class \cite{cambridge2018census}. To effectively assess students and acknowledge prior learning when they switch platforms, smooth "hand-offs" \cite{fancsali2018intelligent} between apps are needed, including a shared taxonomy or translation between taxonomies enabling learning progress on one platform to be continued on the next. Many platforms in the United States have already embarked on migrating to a common taxonomy such as the Common Core State Standards \cite{national2010common} and re-tagging their content or re-mapping their in-house taxonomy to this standard. This task is highly time consuming and likely to be repeated every time a new set of common standards are introduced or modified.

In general, the task of validating the equivalence of skills across taxonomies is non-trivial. For example, the skill of "area of irregular figure" in system A and "area of quadrilaterals and polygons" in system B may or may not be strongly related and it is almost impossible to determine without looking at the set of problems associated with each. 

To date, mapping taxonomies across digital learning platforms using machine learning has largely remained unexplored. They are, however, a prime candidate for this type of approach as they offer data affordances distinct from the standardized testing context, where most taxonomy mapping has occurred. In particular, platforms store data about students’ learning behaviors such as logged clickstream and response sequence data \cite{fischer2020mining}. Some research has successfully harnessed response data to 
infer the underlying skill of problems \cite{pardos2017imputing}.
Here, we evaluate the utility of response sequence data, in addition to problem text, for cross-platform taxonomy mapping.

We show that a taxonomy mapping, or transfer model, can be learned between the skill vector spaces of different platforms, similar to how machine language translation learns mappings between word embedding spaces \cite{mikolov2013exploiting}. Our main contributions are:
\begin{enumerate}
    \item Six models proposed to represent skills as continuous real-valued vectors, with each model able to represent different input data.
    \item Empirical validation of the feasibility of employing machine translation to map between the taxonomies of three digital learning platforms.
    \item Further extension of the methodology to situations where two platforms have asymmetric data types available for representation (e.g., one platform with problem text, the other with only response sequences). 
    \item Inspection of important factors that have an influence on  skill equivalency prediction performance, including model hyperparameters and differences between source and destination taxonomy granularity. 
\end{enumerate}

\section{Related Work}
\label{sec:related-work}
Taxonomy mapping has relied on manual work by subject matter experts \cite{subramaniam2013crosswalk,conley2011crosswalk,razzaq2007developing}, though there has been past research on automating the process using legacy Natural Language Processing (NLP) to find similar skills across taxonomies using text descriptions of each skill \cite{choi2016model,yilmazel2007text}. \citet{choi2016model} converted each skill statement to a verb phrase graph and a noun phrase graph, then calculated similarity between skills by comparing graphs. \citet{yilmazel2007text} used rule-based methods to extract features from skill descriptions in one standard and trained a machine learning classifier to map them to another standard.

Several terms have been used in the literature to refer to the mappings between skill taxonomies. The term "crosswalk" is one, derived from the idea of creating a path to cross a street, used to describe the connection between two taxonomies or sets of educational standards \cite{subramaniam2013crosswalk,conley2011crosswalk}. Other terms like "transfer" \cite{razzaq2007developing} and "alignment" \cite{choi2016model,yilmazel2007text} have also been used in related work. In the context of digital learning environments, several terms have been used to refer to the elements of their taxonomy. Intelligent Tutoring Systems refer to each element as a knowledge component (KC) and a taxonomy of elements as a knowledge component model. A KC is defined as, "an acquired unit of cognitive function or structure that can be inferred from performance on a set of related tasks" \cite{koedinger2012knowledge}. These KCs, tagged to problem steps, allow for adaptive tutoring and cognitive mastery estimation in systems like the Cognitive Tutor (now MATHia) \cite{cen2007over}. The generic term of "skill" \cite{razzaq2007developing} or "tag" \cite{porcello2013crowdsourcing} has been more commonly used in digital learning environments to describe a semi-granular subject area associated with a learning resource. The term "skill" can also be used to generalize the concepts of tags and knowledge components, which is how we will use it throughout the paper. Differences in granularity and epistemology of taxonomies add to the challenge of cross-platform skill equivalency learning.

Methodologically, no prior work has utilized problem text or response sequences, nor have modern neural approaches from computational linguistics been brought to bear on taxonomy mapping. In the broader field of learning analytics research, neural word embeddings \cite{mikolov2013distributed} have been utilized in a number of areas where text descriptions of educational resources are available. Examples include detecting student misconceptions \cite{michalenko2017data}, extracting course concepts \cite{pan2017course}, and coding non-cognitive traits related to student success \cite{stone2019language}. Non-text, sequence data can also carry useful semantic information, such as sequences in which an educational resource or skill appears (i.e., its clickstream logs), which we call "context" information. Previous research has shown that item embeddings learned from sequence contexts with a skip-gram model \cite{mikolov2013distributed} can encode underlying attributes of items helpful to downstream institutional prediction tasks \cite{jiang2020evaluating}, skill label inference \cite{pardos2017imputing}, and course recommendation \cite{morsy2019will,pardos2019connectionist}. After item embeddings are learned, a translation model can be trained to map items from one embedding to another. This is called machine translation and was introduced in the context of language translation using  word embeddings \cite{mikolov2013exploiting,bahdanau2015neural}; however, the idea can be extended to translation between any embedding spaces, such as translating course embeddings between institutions to identify candidate courses for credit articulation \cite{pardos2019data}. 

\section{Models and Methodology}
\label{sec:models}
Each platform consists of problems students are expected to solve that are associated with one or more skills labeled by domain experts. We define the problem of skill equivalency prediction across platforms as follows: given a skill $s$ in a source platform $src$, find $k$ most similar skills ordered by similarity in destination $dst$. The input includes the content and/or context information of skills and some ground truth equivalent skill pairs. More precisely, our method is as follows:
\begin{enumerate}
	\item Represent each skill as a continuous real-valued vector using problem content and/or sequence context.
	\item Learn a translation model from the source skill space to the destination skill space.
	\item Calculate cosine similarities between skills across platforms:
	\begin{equation*}
    \label{eq:cosine}
    \begin{aligned}
      cosine\_similarity(s_{src},s_{dst}) &= \frac{s_{src} \cdot s_{dst}}{\Vert s_{src}\Vert\Vert s_{dst}\Vert} \\ 
      &= \frac{\sum_{i=1}^n s_{src}^i s_{dst}^i}{\sqrt{\sum_{i=1}^n(s_{src}^i)^2}\sqrt{\sum_{i=1}^n(s_{dst}^i)^2}}
    \end{aligned}
    \end{equation*}
	\item For each source skill, rank the destination skills by similarity and take the top $k$ as predictions.
\end{enumerate}

In this section, we describe the models for skill representation and translation. Representation models can be categorized into three types: content-based models (Section \ref{sec:content}), context-based model (Section \ref{sec:context}), and models combining content and context (Section \ref{sec:combined}). The way in which we will translate one representation to another is borrowed from machine translation (Section \ref{sec:machine-translation}). Code to replicate our methodology can be found online\footnote{\url{https://github.com/CAHLR/skill-equivalency}}.

\subsection{Content-based Models}
\label{sec:content}
We represent skills as functions of the problem content they are associated with. The content of a problem in a digital learning platform can include text, graphical figures, and video. For this study, our data only contain the text portion of the content. We utilize the following three content-based representations: Bag-of-words, TF-IDF, and Content2vec.

\subsubsection{Bag-of-words}
Bag-of-words is a standard text processing technique where a document is represented as a vector whose length is equal to the vocabulary size and values are the frequencies of the words occurring in the document. In our experiments, we take all unique words from both platforms as the vocabulary, and represent each problem as a Bag-of-words vector. Then for every skill, we find all problems associated with the skill and average their representations together arriving at a skill vector.

\subsubsection{TF-IDF}
A weakness of Bag-of-words is that the vector might be dominated by frequent but non-distinguishing words. TF-IDF (Term Frequency-Inverse Document Frequency), a method adapted from Bag-of-words, can address this issue, where the values now reflect how distinct a word is to a problem relative to the collection of all problems. The TF-IDF score for word $w$ in problem $p$ from problem set $P$ is calculated by:
\begin{equation}
  TF\text{-}IDF(w,p,P) = TF(w,p) \cdot IDF(w,P)
\end{equation}
where
\begin{equation}
  TF(w,p) = log(1 + freq(w,p))
\end{equation}
\begin{equation}
  IDF(w,P) = log(\frac{|P|}{count(x\in P:w\in x)})
\end{equation}

\subsubsection{Content2vec}
Content2vec is a method built on word embeddings. While previous works have mainly utilized word vectors pretrained on large-scale datasets like Google News corpus \cite{michalenko2017data,stone2019language,pan2017course}, our problem space is different from theirs in that mathematical jargon, symbols, and formulas that are not frequently seen in other corpora take up the majority of our problem texts. Therefore, we test two versions of Content2vec, one with word vectors pretrained on Google News and the other with our own word vectors trained on all the problem texts. We represent each problem as the average of the word vectors, and each skill as the average of the problem vectors.

Note that in content-based methods, representations of problems and skills across platforms share the same space and thus there is no need to translate between spaces for cross-platform skill comparison.

\subsection{Context-based Model (Skill2vec)}
\label{sec:context}
Context information comes from the response log data of a digital learning platform, where each row denotes an interaction between a student and the platform, and pertinent columns include anonymized student ID, attempted problem ID, tagged skills, and start time. To make use of context information we propose Skill2vec. In this method, data are preprocessed into skill sequences per student ordered by start time, and a skip-gram model is then applied to these sequences to learn a continuous vector embedding for each skill. The model is similar to previous work in which problems were embedded based on problem sequence \cite{pardos2017imputing}, but has not until now been applied to skills.

Unlike content-based methods, Skill2vec is trained separately on each platform, hence the generated skill vector spaces are not aligned. Therefore, before measuring cross-platform skill similarity, a translation is necessary to align the two vector spaces. Details are discussed in Section \ref{sec:machine-translation}.

\subsection{Models Combining Content and Context}
\label{sec:combined}
Since both content and context data may contain useful information about skills, we also evaluate the combination of them that may form a more descriptive representation than either alone. In this section, we introduce two models that combine content and context information: a simple concatenation of our previously mentioned content and context vectors and a separate model from the literature, called Text-Associated Matrix Factorization (TAMF), that integrates both during training.
\subsubsection{Content2vec+Skill2vec}
\begin{figure}[htb]
    \centering 
    \includegraphics[width=\linewidth]{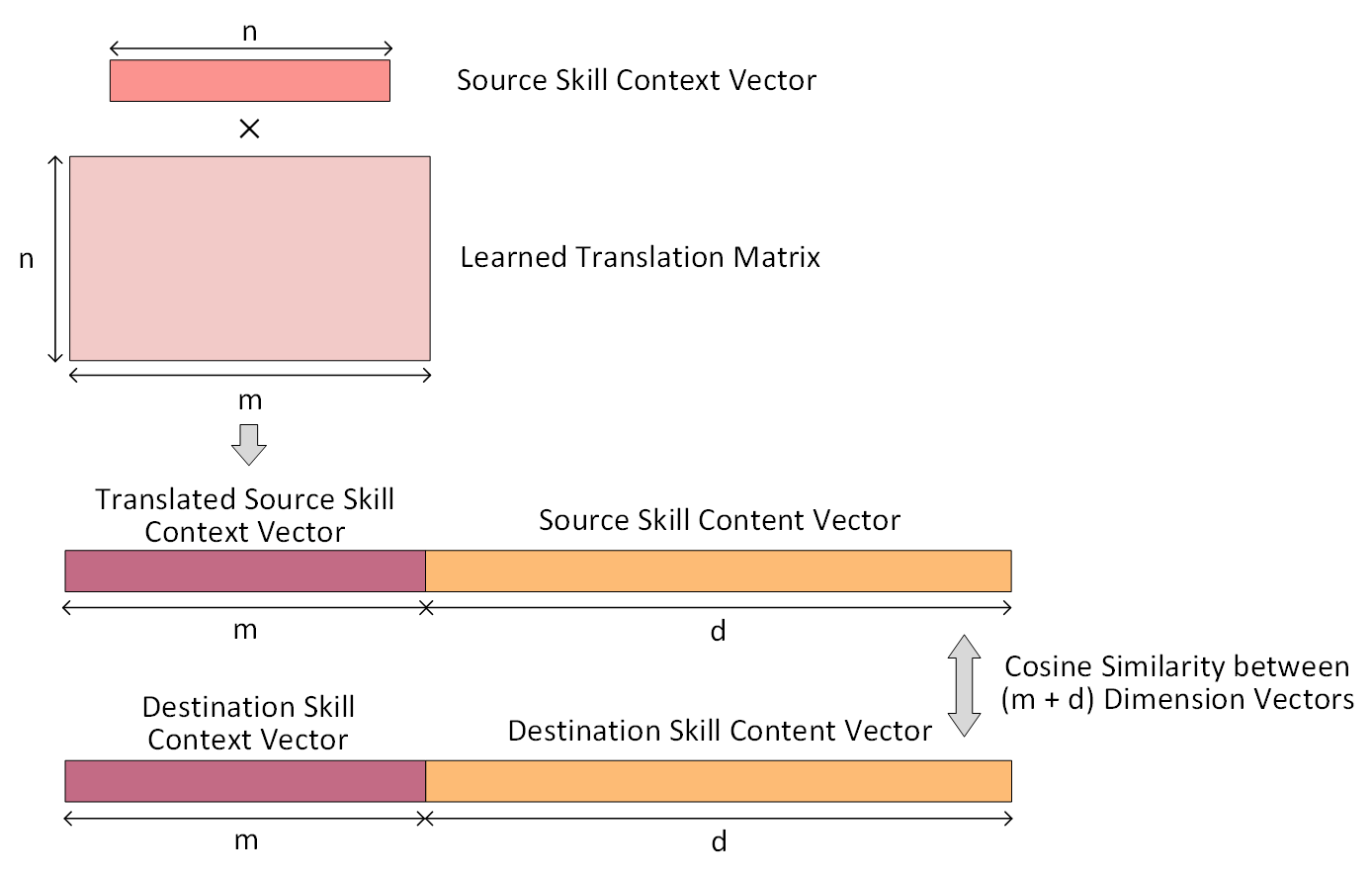}
    \caption{Process of combining Skill2vec and Content2vec representations}
    \label{fig:combined}
    \Description{A diagram showing how to combine Skill2vec and Content2vec}
\end{figure}

The first model is simply the concatenation of Content2vec and Skill2vec. Since Content2vec does not need translation but Skill2vec does, we design the process of concatenation as follows (Figure \ref{fig:combined}): we first learn a translation model for context vectors, and then combine the translated source context vector with the source content vector, while the destination skill vector is obtained by appending the destination content vector to the unchanged destination context vector. By doing so, we make sure the two components reside in the same space, thus allowing the concatenated vectors to be compared.

\subsubsection{Text-Associated  Matrix Factorization}
While concatenation of representations is commonplace, we also sought out a method that learns a single representation from both sources. Text-Associated Matrix Factorization (TAMF) \cite{yang2015network} is such a method and may be able to learn features related to the interaction between content and context, and could in principle be more expressive than simple concatenation where no interaction between sources can be utilized.

The method learns embeddings based on nodes in a graph and incorporates content information in a matrix factorization stage. Our adaptation differs from the original model in that they begin with a graph and use Deepwalk \cite{perozzi2014deepwalk} to generate a context matrix from which item embeddings are learned, while our response log data are already sequences, therefore we factorize the Positive Pointwise Mutual Information (PPMI) matrix derived from our skill sequences instead of a Deepwalk matrix.

Specifically, let $S$ be the set of all skills and $P$ be the set of all skill and context skill pairs observed in the input data, then the PPMI matrix $M\in\mathbb{R}^{|S|\times|S|}$ can be calculated by $M_{s,c}=\max(0, \log\frac{\#(s,c)\cdot|P|}{\#(s)\cdot\#(c)})$ where $s$ is a skill, $c$ is a context skill, and $\#(s)$, $\#(c)$, and $\#(s,c)$ are the counts of occurrences of $s$, $c$, and the pair $s,c$ within a window size in all sequences. According to \citet{levy2014neural}, a skip-gram model (Skill2vec in our case) implicitly factorizes the PPMI matrix $M=W^\top H$ where $W,H\in\mathbb{R}^{k\times|S|}$, with $k$ being the dimension of the learned embeddings. The matrix $W$ is equivalent to the output embeddings of the skip-gram model.

\begin{figure}[htb]
    \centering 
    \includegraphics[width=\linewidth]{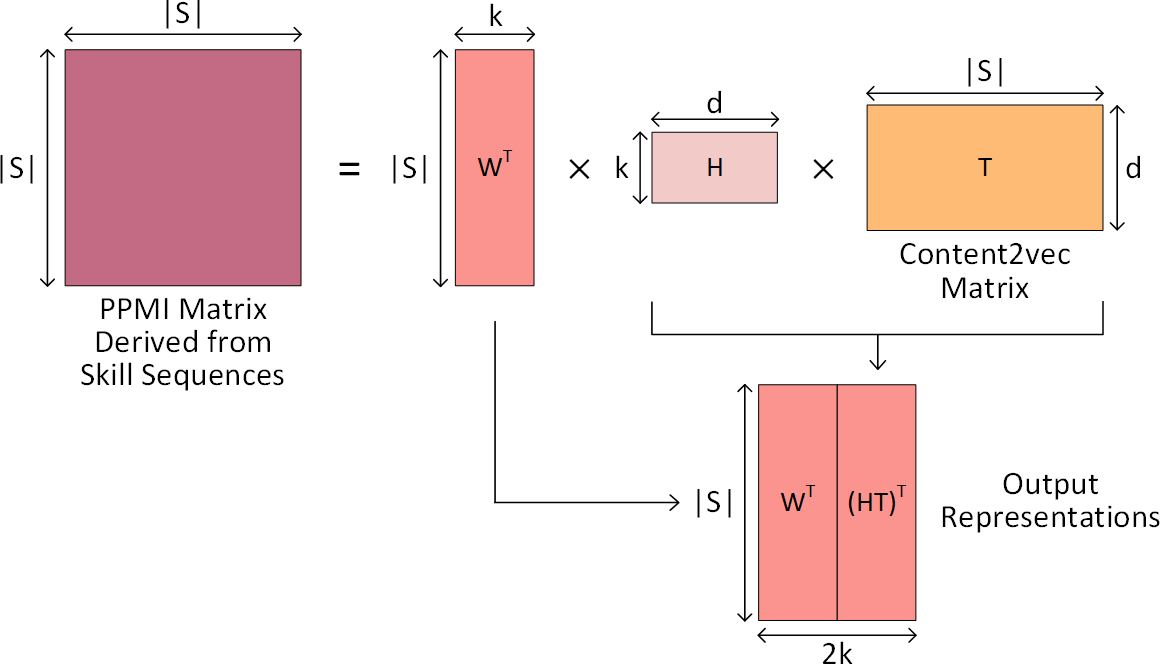}
    \caption{Text-Associated Matrix Factorization (TAMF)}
    \Description{A diagram of matrix factorization}
    \label{fig:tamf}
\end{figure}

With TAMF, the problem is formulated as $M=W^\top HT$, where $W\in\mathbb{R}^{k\times|S|}$, $H\in\mathbb{R}^{k\times d}$, and  $T\in\mathbb{R}^{d\times|S|}$ (Figure \ref{fig:tamf}). The new matrix $T$ is the Content2vec matrix of embedding size $d$. The output of the method is the concatenation $[W^\top, (HT)^\top]\in\mathbb{R}^{|S|\times 2k}$ with embedding size $2k$. In this way content and context are deeply merged in the skill representations.

The optimization objective of TAMF is to search for $W$ and $H$ that minimizes the loss:
\begin{equation*}
L=\Vert M-W^\top HT\Vert^2_F + \frac{\lambda}{2}(\Vert W\Vert^2_F+\Vert H\Vert^2_F)
\end{equation*}

where the first term $\Vert M-W^\top HT\Vert^2_F$ is the distance between the original PPMI matrix and the reconstructed matrix from the learned factorization, and the second term $\frac{\lambda}{2}(\Vert W\Vert^2_F+\Vert H\Vert^2_F)$ is a regularization term restricting the magnitude of the learned matrix. Minimizing this loss should give us a reasonable factorization while not overfitting to the data.

This loss $L$ is convex with respect to $W$ when $H$ is fixed and vice versa, hence we can iteratively optimize in closed form by taking partial derivatives, reorganizing, and solving linear systems for $W$ and $H$  given the other one fixed until the decreasing rate of $L$ is below a certain threshold. While this process does not guarantee to hit the global minimum, in practice we find it works well.\\

\subsection{Machine Translation}
\label{sec:machine-translation}
For Skill2vec and TAMF, representations of skills in different platforms are learned independently and do not share the same space. They may not even have the same dimensionality. To align vector spaces, we use machine translation to learn a transformation from a source skill vector space to destination space. Previous research has found that linear transformation outperforms more complex neural translation models on word-to-word translation tasks \cite{mikolov2013exploiting}. Therefore we choose to learn a linear translation matrix $T\in\mathbb{R}^{m\times n}$ that maps a source skill vector $v_{s}\in\mathbb{R}^n$ to a vector $v_{d}\in\mathbb{R}^m$ in the destination vector space by maximizing the cosine similarities of translated source vectors and ground truth destination vectors based on a set of known equivalent skill pairs.

\begin{table*}
\caption{Descriptive statistics of datasets (after preprocessing)}
\label{tab:datasets}
\begin{tabular}{cccc}
\toprule
& ASSISTments & Khan Academy &  Cognitive Tutor\\
\midrule
         Subject &Math &Math &Math \\
         Grade Level &7th and 8th grades &7th and 8th grades &8th to 10th grades\\
         Number of Skills   &130     &194     &536\\
         Number of Unique Problems (Steps for Cognitive Tutor) &38,490   &20,797 & 495,068\\
         Number of Unique Users    &27,760   &875,492   &3,292\\
         Number of Responses     &2,602,777 &47,794,008  &6,263,006\\
         Average Number of Words per Problem &32 &29 & N/A \\
         Information Types & Content+Context & Content+Context & Context\\
         Granularity & Coarse-grained & Medium & Fine-grained\\
\bottomrule
\end{tabular}
\end{table*}

\section{Experimental Setup}
\label{sec:experimental-setup}
\subsection{Datasets}
The datasets for this study are from three digital learning platforms: ASSISTments\footnote{\url{https://www.assistments.org}}, Khan Academy\footnote{\url{https://www.khanacademy.org}}, and Cognitive Tutor (now named MATHia\footnote{\url{https://www.carnegielearning.com}}). All three platforms offer content primarily for middle school and high school students.
\subsubsection*{ASSISTments}
We use the public ASSISTments 2012-2013 dataset\footnote{\url{https://sites.google.com/site/assistmentsdata/home/2012-13-school-data-with-affect}} including problem texts, which contains problems for 7th and 8th grades math \cite{heffernan2014assistments}. Most problems in ASSISTments are tagged with a single skill and we keep only these problems and keep only skills with at least 1,000 response logs associated with them. We apply this filter since ASSISTments contains much content that is teacher produced for a single class. ASSISTments has the most coarse-grained taxonomy among the three platforms.
\subsubsection*{Khan Academy}
We use anonymized student responses collected from Khan Academy 7th and 8th grades math exercises between 2013-2014, and collect problem texts through web scraping. Because there are no explicit skills assigned in our data, and each exercise in Khan Academy serves as a template generating multiple problems, we decide to regard each exercise as a skill, as was done in \citet{piech2015deep}. This results in a very minor difference between some skills, (e.g., $combining\_like\_terms\_1$ and $combining\_like\_terms\_2$ are considered as two skills). We find that keeping these skills separate gives better performance than grouping them together. Khan Academy's taxonomy is finer than ASSISTments, but coarser than Cognitive Tutor.
\subsubsection*{Cognitive Tutor}
We use the publicly available Cognitive Tutor dataset from the 2010 KDD Cup \cite{stamper20162010}. We choose the Algebra I 2008-2009 challenge dataset as it is the largest and its content covers 8th-10th grade math \cite{pane2014effectiveness} with good overlap with the other two platforms in our study. Cognitive Tutor skills are assigned per step, rather than per problem, and it is not uncommon for each step to have more than one skill. To allow for multiple skills, we generate the skill sequences for Skill2vec by randomly ordering skills within a single step. We select the column "KC(SubSkills)" to serve as the skills column since it has the least number of missing values compared to alternative KC columns. Note that problem texts (i.e. content information) are not present in Cognitive Tutor, raising the issue of asymmetric data addressed in Section \ref{section:results:various-source-destination}. The taxonomy of Cognitive Tutor is the finest-grained.
\subsubsection*{Preprocessing}
For all three datasets, we drop the rows without skill assignment. For problem texts we tokenize words and clean the texts by removing stop words and converting to lower case. The basic descriptives of the datasets after preprocessing are shown in Table \ref{tab:datasets}.
\subsubsection*{Skill Equivalence Labeling}
There are currently no skill equivalency labels across these three platforms, so we create the labels ourselves. For each pair of platforms, the three authors separately annotated equivalent skills across platforms and then agreement was measured by Fleiss' Kappa and any conflict was resolved by discussion and majority voting.  To give an example, the authors initially disagreed on whether skill "Angles 2" in Khan Academy should be mapped to skill "Angles - Obtuse, Acute, and Right" in ASSISTments. After discussion and inspecting the associated problems, a consensus was arrived at that they should not be linked as the Khan Academy skill involves angle calculation which is not needed in the ASSISTments skill. Since the granularity of taxonomies varies, there is likely to be one-to-many and many-to-one relationships between taxonomies. To account for such asymmetric relationships, we calculate Fleiss' Kappa by considering every possible pair of skills as "subject", and a binary value indicating equivalent or not as label. The Fleiss' Kappas and number of equivalent skill pairs are displayed in Table \ref{tab:kappa}. The high Kappas suggest a good agreement among the authors.

\begin{table}
\small
\caption{Annotation results}
\label{tab:kappa}
\begin{tabular}{ccc}
\toprule
\multirow{2}{*}{Platforms} & \multirow{2}{*}{Fleiss' Kappa} & Number of Skill \\
&&Equivalencies \\

\midrule
         Khan Academy, ASSISTments & 0.69 &148\\
         Khan Academy, Cognitive Tutor  & 0.72 &85\\
         ASSISTments, Cognitive Tutor  & 0.83 &222\\
\bottomrule
\end{tabular}
\end{table}

\subsection{Evaluation Metrics}
We evaluate our models using two metrics, recall@$k$ and mean reciprocal rank. The skill taxonomies in our platform datasets differ in granularity, resulting in some skills having a one-to-many relationship across taxonomies. Thus, in addition to using a somewhat standard metric of recall@$k$ \cite{pardos2019data}, we also report mean reciprocal rank to indicate the rank of the first true positive result.

Recall@$k$, where $k$ is a user definable integer, measures the percentage of relevant destination skills that are contained within the model's top $k$ predictions for each source skill, as calculated by $\frac{true\ positives@k}{true\ positives@k + false\ negatives@k}$. In our experiments, we choose $k$ to be 5, which will return the 5 most similar predicted skills.

Mean reciprocal rank (MRR) calculates the rank of the first relevant destination skill for each source skill and takes the mean of the reciprocals of these ranks as output. It is given by
  $\frac{1}{|S|}\sum_{s\in S}\frac{1}{rank(s)}$
%\end{equation}
where $S$ is the set of all source skills and $rank(s)$ denotes the rank of the first relevant destination skill for source skill $s$.

\subsection{Experiment Details}
\label{experimental-setup:implementation}

\subsubsection{Validation and Testing}
\label{sec:validation}
We take different validation and testing strategies for different representation models, depending on whether they require hyperparameter tuning and whether they need ground truth labels to train. We apply hyperparameter search on validation data separate from the test set.

For Bag-of-words, TF-IDF, and Content2vec with pretrained word vectors, no hyperparameters or ground truth labels are involved in training. Test set results are produced by comparing the predictions against all the ground truth pairs.

For Content2vec with our own word vectors trained on problem texts, no skill pair labels are needed, but the word vector training process involves hyperparameters. Using 10-fold cross-validation, we search the best hyperparameters on the training set of each phase of the cross-validation and evaluate on the phase's test set with those hyperparameters, hence we get 10 best hyperparameter sets for the 10 folds instead of a single best hyperparameter set.

For Skill2vec and TAMF, the training requires both hyperparameter tuning and ground truth skill pairs. We similarly conduct a nested cross-validation on the training data within each phase of the 10-fold cross-validation to tune hyperparameters. The outer cross-validation tier has the same splits with Content2vec in order to compare models.

For Content2vec+Skill2vec, we simply concatenate the tuned content vector and context vector with the best hyperparameters for each fold.

At the dataset level, we treat Khan Academy and ASSISTments as training platforms and run experiments on them to compare models. We then pick the best model and conduct a final test on the mappings to and from Cognitive Tutor. We choose Cognitive Tutor as the test platform because it has only context information, whereas ASSISTments and Khan Academy have both content and context with which we can do holistic experiments on various skill representations, including symmetric and asymmetric source and destination data sources. A test on Cognitive Tutor without any model tuning will validate (or invalidate) our conclusions drawn from the other two platforms.

\subsubsection{Training}
We use Adam optimizer with learning rate 0.001 to train the translation model. The loss to minimize is the mean of cosine distances of training ground truth skill pairs. The maximum number of epochs is 1,000. We also randomly take 20\% out of the training data as validation. If the validation loss has not decreased in 100 epochs, we stop the training and keep the model with the smallest validation loss. The models are trained on one NVIDIA Maxwell architecture GPU and each training job takes around 10 seconds to finish, with a whole nested cross-validation with 24 hyperparameter sets to tune totaling 5 hours ($10 \text{ seconds} \times 10 \times 8 \times 24 = 19200$ seconds) to run.

\section{Results}
\label{sec:results}
This section presents the results of several experiments. Section \ref{section:results:models} compares equivalent skill prediction performance of the six skill representations in Khan Academy and ASSISTments assuming the same source and destination information types. Section \ref{section:results:various-source-destination} examines situations where source and destination platforms have different types of skill information (e.g., content in one and context in the other). Section \ref{section:results:cog} evaluates skill prediction performance on Cognitive Tutor as a test set using the best models as observed from the Khan Academy and ASSISTments results. Section \ref{section:results:hyperparameter} introduces the ablation study results for each method and the performance variations with different hyperparameters. 

\subsection{Symmetric Data Sources}
\label{section:results:models}

\begin{table}
  \caption{Skill equivalency prediction results using symmetric data sources and skill representations}
  \label{tab:models}
  \begin{tabular}{ccccc}
    \toprule
        \multirow{2}{*}{Representation} &K2A &K2A &A2K &A2K\\
        &Recall@5 &MRR &Recall@5 &MRR\\
        % Representation & K2A Recall@5 & K2A MRR& A2K Recall@5& A2K MRR \\
    \midrule
     Bag-of-words   &0.55 &0.43 &0.38 &0.46 \\
     TF-IDF           &0.51 &0.44 &0.43 &0.51 \\
     Content2vec (pretrained)   &0.33 &0.21 &0.14 &0.25 \\
     Content2vec    &0.64 &0.49 &0.44 &0.55 \\
     Skill2vec    &0.61 &0.51 &0.11 &0.14 \\
     Content2vec+Skill2vec    &0.80 &0.63 &0.28 &0.35 \\
     TAMF    &0.74 &0.67 &0.22 &0.21 \\
  \bottomrule
\end{tabular}
\end{table}

In this section, we assume the source and destination platforms have the same type of data available (content, context, or both), and compare the six skill representations detailed in Section \ref{sec:models} on the skill equivalency prediction tasks between Khan Academy and ASSISTments. The results are shown in Table \ref{tab:models}, where "K2A" refers to mapping from Khan Academy to ASSISTments taxonomy, and "A2K" is the other direction. The main observations are as follows:
\begin{itemize}
\item The highest MRRs for both directions are above 0.5, meaning the true predictions are at the second place or higher on average.
\item Among all content-based methods, Content2vec with word vectors trained on problem texts is consistently the best. Content2vec with pretrained word vectors perform poorly, perhaps due to the insufficiency of the pretrained corpus in capturing information on math problem texts.
\item In the mapping direction from Khan Academy to ASSISTments, incorporating both content and context information in the representation is better than either alone. The best method depends on the evaluation metric: Content2vec+Skill2vec is preferred with recall@5, and TAMF is favorable with MRR.
\item In the mapping direction from ASSISTments to Khan Academy, the best method is Content2vec. Moreover, sequence information is counterproductive since Content2vec+Skill2vec is worse than Content2vec.
\item The direction from Khan Academy to ASSISTments has better results than the other direction. This is likely because Khan Academy has a finer-grained taxonomy than ASSISTments, which will be discussed further in Section \ref{section:analysis:discrepancy}.
\end{itemize}

\subsection{Asymmetric Data Sources}
\label{section:results:various-source-destination}
In certain scenarios, the combination of content and context information for a digital learning platform will not be available. For example, Cognitive Tutor provides only context information publicly, and other platforms like Junyi Academy\footnote{\url{https://github.com/junyiacademy/junyiexercise}} have only content information made available. While the platforms themselves likely store sequence data, a new platform with little to no sequence data may still wish to map to other taxonomies. Our framework can accommodate such circumstances: it allows different types of source and destination skill representations to be presented to the model. The machine translation will learn transformations between the two vector spaces provided that there are ground truth skill pairs to train on. Experiments are conducted to test how skill equivalency prediction performance differs when data type availability is not symmetric between source and destination. There are three types of information input for a given platform: content, context, and both, resulting in 6 different combinations of asymmetric source and destination representation types. Table \ref{tab:results:src-dest} gives the results of the different representation scenarios. The best models in terms of MRR for each input type (Content2vec, Skill2vec, and TAMF) are used.

\begin{table*}
  \caption{Skill equivalency prediction results using asymmetric data sources and skill representations}
  \label{tab:results:src-dest}
  \begin{tabular}{cccccccccc}
    \toprule
        Source &Source&Source& Destination& Destination& Destination & K2A & K2A & A2K & A2K \\
        Content&Context&Model& Content&Context &Model&Recall@5 & MRR & Recall@5 & MRR\\
    \midrule
    %  \checkmark & & Content2vec & \checkmark & & Content2vec &0.64 &0.49 &\bf{0.44} &\bf{0.55} \\
     \checkmark & & Content2vec & & \checkmark & Skill2vec &0.71 &0.57 &0.07 &0.13 \\
     \checkmark & & Content2vec & \checkmark & \checkmark & TAMF &0.71 &0.55 &0.19 &0.18 \\
     & \checkmark & Skill2vec & \checkmark & & Content2vec & 0.47 &0.35 &0.11 &0.17 \\
    %  & \checkmark & Skill2vec & & \checkmark & Skill2vec &0.61 &0.51 &0.11 &0.14 \\
     & \checkmark & Skill2vec & \checkmark & \checkmark & TAMF & 0.69 &0.54 &0.15 &0.14 \\
     \checkmark & \checkmark & TAMF & \checkmark & & Content2vec &0.69 &0.60 &0.24 &0.22 \\
     \checkmark & \checkmark & TAMF & & \checkmark & Skill2vec &0.70 &0.63 &0.12 &0.16 \\
    %  \checkmark & \checkmark & TAMF & \checkmark & \checkmark & TAMF &\bf{0.74} &\bf{0.67} &0.22 &0.21 \\
  \bottomrule
  \end{tabular}
\end{table*}
Comparing the results with Table \ref{tab:models}, we see that the highest recall@5 and MRR with asymmetric data sources are lower than symmetric cases, so it's better to first consider symmetric sources if all information is available. However, asymmetric sources are still a viable option. On one hand, asymmetric sources can not be avoided if one platform has only content and the other has only context. Moreover, even if one platform has both types of information and the other has only content or context, where training with symmetric sources is an option, it's still worth trying to incorporate both information in the first platform. For example, as shown in Table \ref{tab:models} and \ref{tab:results:src-dest}, TAMF to Skill2vec (asymmetric) is better than Skill2vec to Skill2vec (symmetric).

\subsection{Testing on Cognitive Tutor}
\label{section:results:cog}
We choose the models for mapping between Cognitive Tutor and the other two platforms anticipated to be the best based on the results of mapping shown in Tables \ref{tab:models} and \ref{tab:results:src-dest}. For each mapping direction, given that Cognitive Tutor has only context data, we look for the best representation for the other platform. For example, when selecting for Cognitive Tutor to Khan Academy, we look at the results where Khan Academy is the destination and the source is context only, and find the content-only destination has the highest MRR. We therefore choose Content2vec for Khan Academy.

The skill equivalency prediction results are displayed in Table \ref{tab:cog}. Mapping from Cognitive Tutor to the other two platforms reach relatively high recall@5 and MRR, supporting our methodology and conclusions from experiments between Khan Academy and ASSISTments; however, the opposite directions are worse. We believe this is because Cognitive Tutor has the finest-grained skill taxonomy and naunced information is lost when forcing a mapping from coarse-grained to fine-grained (i.e., one-to-many).

\begin{table*}
  \caption{Skill equivalency prediction results on Cognitive Tutor}
  \label{tab:cog}
  \begin{tabular}{cccccc}
    \toprule
        Source&Source Representation &Destination&Destination Representation &Recall@5 &MRR\\
    \midrule
     Cognitive Tutor &Skill2vec  & Khan Academy   & Content2vec  &0.72  &0.72 \\
     Khan Academy&TAMF  & Cognitive Tutor   & Skill2vec  &0.23   &0.39 \\
     Cognitive Tutor &Skill2vec &ASSISTments   &TAMF  &0.88  &0.79 \\
     ASSISTments   &TAMF     &Cognitive Tutor &Skill2vec  &0.03  &0.12 \\
  \bottomrule
\end{tabular}
\end{table*}

\subsection{Hyperparameter Tuning}
\label{section:results:hyperparameter}
Hyperparameters are tuned in the experiments with symmetric data sources. As detailed in Section \ref{sec:validation}, three methods need hyperparameter tuning: Content2vec, Skill2vec, and TAMF. Although we do not find a single best hyperparameter set as a consequence of cross-validation, we can compare the average performance of each hyperparameter set on the validation set. The metric is MRR in hyperparameter tuning thanks to its insensitivity to the one-to-many relationship.

For Content2vec, the hyperparameters include vector dimension, window size, and minimum count. Surprisingly, the best hyperparameters for each mapping direction are the same across all 10 folds, which are vector dimension 100, window size 20, and minimum count 50 from Khan Academy to ASSISTments, and vector dimension 100, window size 20, and minimum count 30 from ASSISTments to Khan Academy. Generally speaking, larger vector dimension and larger window size are better, while minimum token count does not have much impact on the performance.

\begin{figure}[htb]
  \centering
  \includegraphics[width=\linewidth]{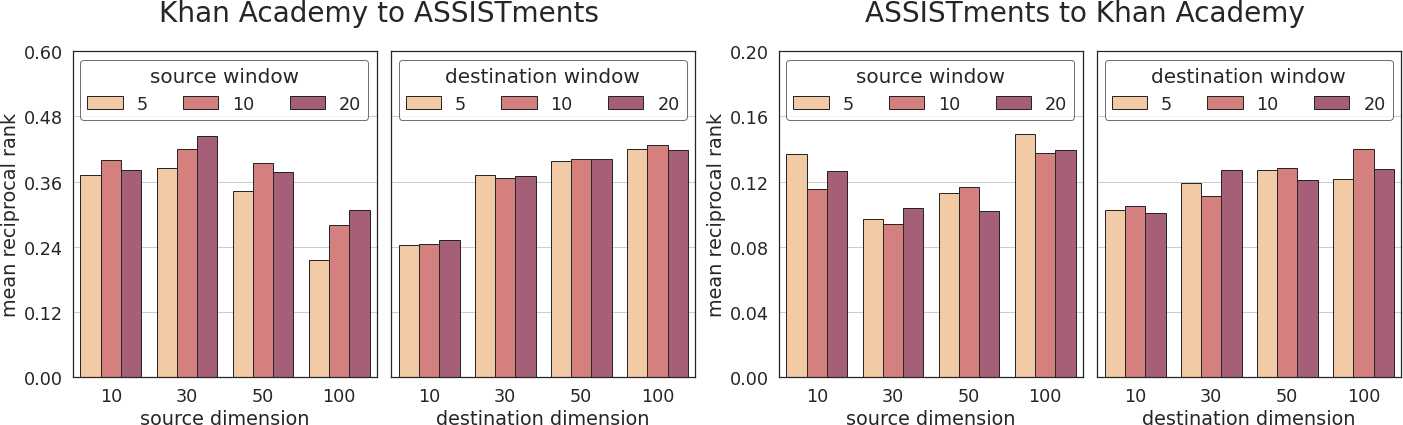}
  \caption{Skill2vec performance with different hyperparameters}
  \Description{A barplot showing the average performance of different hyperparameter sets in Skill2vec}
  \label{fig:results:hyperparamter:skill2vec}
\end{figure}

For Skill2vec, the source and destination hyperparameters can be different, but there is no minimum count as we need to keep all skills, so the hyperparameters are source vector dimension, source window size, destination vector dimension, and destination window size. Figure \ref{fig:results:hyperparamter:skill2vec} shows how they affect validation MRR across all folds. We can see that vector dimension influences the results more significantly than window size, but there is no single rule to choose vector dimension, since sometimes large vector dimension is favorable while sometimes small dimension is better.

\begin{figure}[htb]
  \centering
  \includegraphics[width=\linewidth]{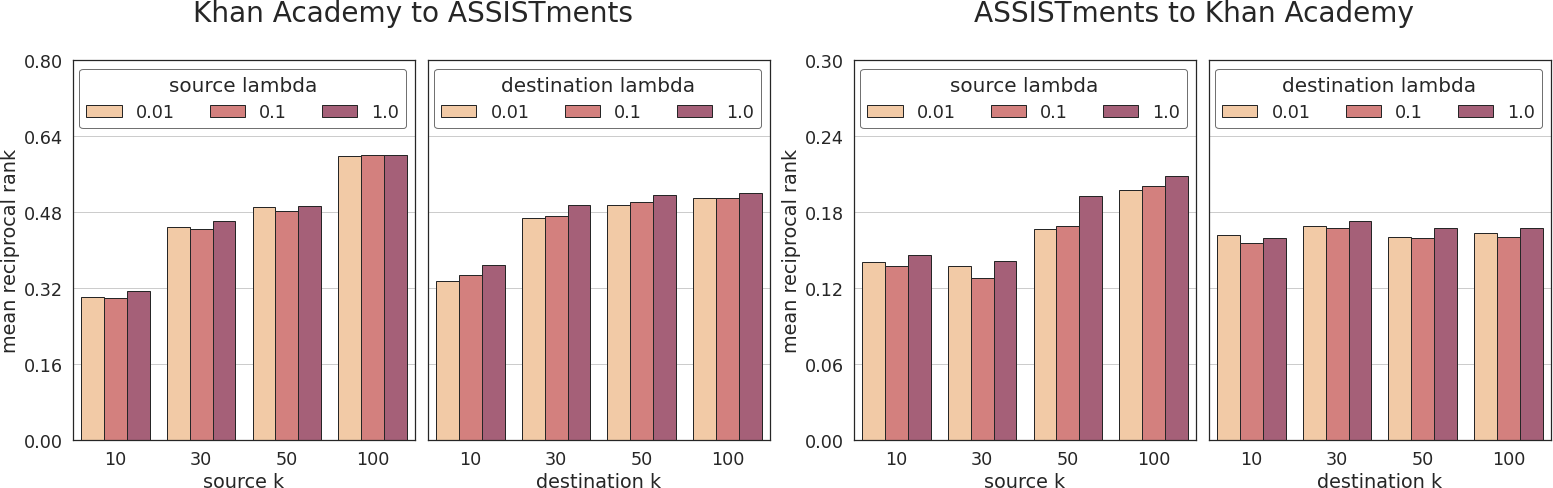}
  \caption{TAMF performance with different hyperparameters}
  \Description{A barplot showing the average performance of different hyperparameter sets in TAMF}
  \label{fig:results:hyperparamter:TAMF}
\end{figure}

TAMF has two hyperparameters for each platform, $k$ for half vector dimension and $\lambda$ for regularization coefficient. Shown in Figure \ref{fig:results:hyperparamter:TAMF} is the performance with different hyperparameter sets. Larger vector dimension is preferred, and regularization coefficient does not greatly affect the overall skill equivalency prediction results.

\section{Analysis}
\label{sec:anslysis}
This section delves into inspecting the taxonomy translation and edge cases. Section \ref{section:analysis:example} visualizes skills from all taxonomies mapped to a single shared space; Section \ref{section:analysis:counterpart} considers skills with no matching skills in the destination platform; and Section \ref{section:analysis:discrepancy} summarizes the effect of taxonomy granularity on skill equivalency prediction.

\subsection{Visualization of Mapped Skills}
\label{section:analysis:example}
\begin{figure*}[h]
    \centering 
    \includegraphics[width=\linewidth]{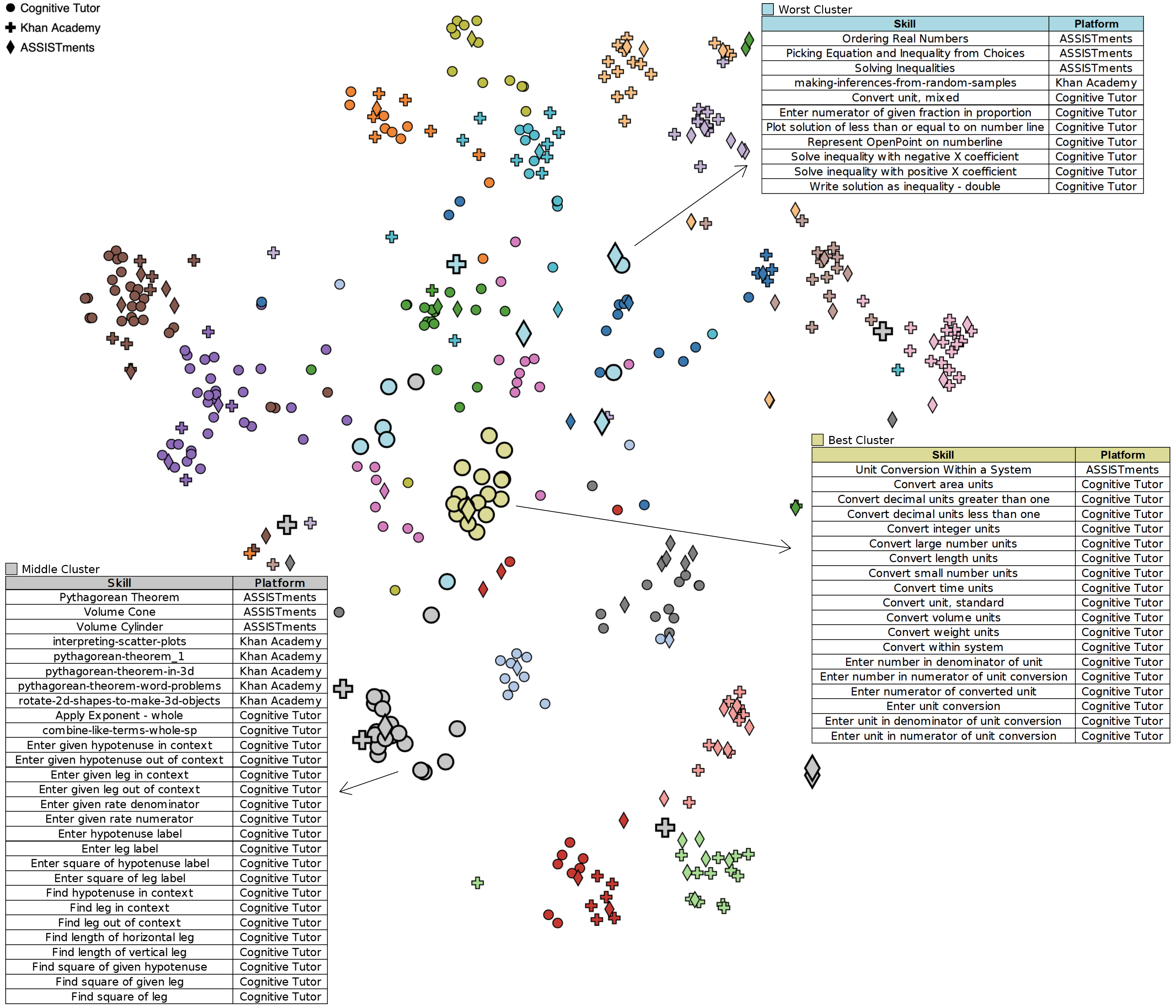}
    \caption{t-SNE visualization of skills from all three platform taxonomies projected onto the ASSISTments TAMF vector space}
    \Description{A scatter plot with skills as points shaped by platform and colored by cluster. Three clusters are highlighted and the skills are listed in tables.}
    \label{fig:examples}
\end{figure*}

We map skills from all platform taxonomies to a common space and visualize them to help understand skill equivalency in our taxonomy mapping models. Specifically, we project Khan Academy TAMF vectors and Cognitive Tutor Skill2vec vectors onto the ASSISTments TAMF vector space and keep ASSISTments TAMF vectors unchanged, since the skill mapping in this direction gives the best equivalency prediction performance. In this space, we run k-means clustering to explore relationships among closely mapped skills. The number of clusters is set to 20 as determined by the "elbow method" heuristic. Finally, we apply t-SNE to reduce dimensionality and display the results in Figure \ref{fig:examples}. Skills are colored based on k-means applied to the original, high dimensional skill vectors.

We rank these clusters by a heuristic score: the percent of skills whose true matching skills from another taxonomy are also in the same cluster. The best cluster (score 1.0), the worst cluster (0.36), and a middle cluster (0.71) are featured in Figure \ref{fig:examples}. The best cluster is tightly grouped, containing only unit conversion skills. The worst cluster is the most dispersed, yet it still groups similar skills together like inequality and number line. The middle cluster is mostly about Pythagorean Theorem, with a few outliers. In this cluster, we can observe the fined-grained step-level skills of Pythagorean Theorem like calculating lengths and squares of hypotenuse and legs from Cognitive Tutor are clustered together with the coarse-grained problem-level skills from the other two platforms. The clustering results indicate that our models do map similar skills close to each other as desired. 
The complete visualization with cluster assignments and plot inspection tool can be found online\footnote{ \url{https://cahlr.github.io/skill-equivalency-visualization}}.

\subsection{Skills without Counterparts}
\label{section:analysis:counterpart}
Owing to the uniqueness of each learning platform, some skills in one platform may not have a matching skill in another. In our case, 30\% of skills in Khan Academy have no corresponding skills in ASSISTments and 31\% of skills in ASSISTments have no counterparts in Khan Academy. We call these skills the "None" skills and they were ignored in the previous evaluations. While an algorithm can always return the most similar but not sufficiently equivalent skills, it is desirable to find a way for our method to report that there is no reasonable match. Therefore, we also conduct experiments to distinguish these untranslatable skills. Our method is to simply add a similarity threshold, whereby any prediction below that threshold will be considered as a "None" skill prediction, and the ground truth for those untranslatable skills is a mapping to the "None" skill. The evaluation metric used is again recall@5. We also compare this method against a baseline "Ignore", which is to train and predict as if all skills are translatable but consider the predictions for those "None" skills as wrongly predicted in evaluation.

\begin{table}
  \caption{Skill equivalency prediction results including "None" skills}
  \label{tab:none}
  \begin{tabular}{cccc}
    \toprule
        \multirow{2}{*}{Representation} &"None" Skill &K2A &A2K\\
        &Strategy & Recall@5& Recall@5\\
    \midrule
     Content2vec & Ignore   &0.45 &0.30 \\
     Content2vec & Threshold   &0.53 &0.42 \\
     Skill2vec  & Ignore  &0.43 &0.08\\
     Skill2vec  & Threshold  &0.47 &0.13\\
     Content2vec+Skill2vec  & Ignore  &0.56 &0.24\\
     Content2vec+Skill2vec  & Threshold  &0.62 &0.26\\
  \bottomrule
\end{tabular}
\end{table}

Table \ref{tab:none} shows the results for Content2vec, Skill2vec, and Content2vec+Skill2vec, which are the best method with recall@5 for each input type. Considering "None" skills leads to a decrease in performance compared to previously shown evaluations, underscoring the difficulty of addressing this issue. The threshold approach, however,  performs slightly better than the baseline of not modeling None skills and counting them wrong. Given the simplicity of the threshold method, we hope future research can improve on this approach to classifying untranslatable skills between platforms.

\subsection{Impact of Taxonomy Granularity on Skill Equivalency Prediction}
\label{section:analysis:discrepancy}
We observed that there is a large performance discrepancy between directions of taxonomy mapping. For example, the best MRR from Khan Academy to ASSISTments is 0.67 achieved with TAMF, but the best MRR in the other direction is only 0.55. This phenomenon is also observed in other platform pairs (Table \ref{tab:discrepancy}). For example, results of Cognitive Tutor to Khan Academy or ASSISTments are better than Khan Academy or ASSISTments to Cognitive Tutor.

A common characteristic of the poorer performing pairs is when mapping in the direction of a fine-grained taxonomy to a coarse-grained one. Cognitive Tutor is the finest with skills assigned to step, Khan Academy is the middle, and ASSISTments uses the coarsest-grained taxonomy of the three. We can observe the magnitude of the difference in performance between directions differs, as shown in Table \ref{tab:discrepancy}. Cognitive Tutor and ASSISTments have larger difference than Cognitive Tutor and Khan Academy, and Khan Academy and ASSISTments. This result suggests that the more distant two platforms are in granularity, the larger performance discrepancy there will be between the two mapping directions.

\begin{table*}
  \caption{Discrepancy of skill equivalency prediction results (MRR) between opposite directions}
  \label{tab:discrepancy}
  \begin{tabular}{ccccc}
    \toprule
        Fine-grained Platform & Coarse-grained Platform & Fine to Coarse  & Coarse to Fine & Difference\\
    \midrule
     Khan Academy & ASSISTments   &0.67  &0.55   &0.12 \\
     Cognitive Tutor & Khan Academy  &0.72  &0.39  &0.33 \\
     Cognitive Tutor & ASSISTments  &0.79 &0.12  &0.67 \\
  \bottomrule
\end{tabular}
\end{table*}

\section{Limitations and Future Work}
\label{section:analysis:limitation}
There were a few limitations to this research that we believe can be improved in future studies. First, the ground truth cross-platform skills were labeled by the authors, not domain experts. Second, the taxonomies used are restricted within the mathematics domain. This was in part because the majority of publicly available digital learning platform clickstream and content datasets are from mathematics platforms. Third, we only utilized text content in our content-based approaches. Utilizing images, as was done in \cite{chaplot2018learning}, would capture a more complete representation of problems. Finally, pretrained contextual word embedding models like BERT \cite{devlin2019bert} might be effective in improving content-based representations of problems, thus boosting skill equivalency prediction performance.

\section{Conclusion}
In this research, we demonstrated the viability of learning skill equivalencies across several taxonomies using data from the content of the problems skills are associated with and the clickstream sequences (i.e., contexts) in which those problems appear on a digital learning platform. %We find that both content and context information are useful in learning skill representations to predict cross-platform equivalencies, but the best representation and hyperparameters vary with platforms, transfer directions, and evaluation metrics. 
We represented skills as vectors and employed machine translation to map between skill vector spaces, and validated the methodology on three digital learning platforms: ASSISTments, Khan Academy, and Cognitive Tutor. 

We found that attempting to map from a coarser-grained taxonomy to a finer-grained taxonomy was considerably more difficult, with the best ASSISTments to Khan Academy recall@5 markedly lower than the reverse direction (0.44 vs 0.80) and, similarly, when mapping to the finer-grained taxonomy of the Cognitive Tutor (0.23 and 0.03 with Khan Academy and ASSISTments as sources, respectively). We also found that skill equivalence prediction was more accurate in experiments where there was symmetric data used to represent skills in both the source and destination taxonomies; however, skill prediction with asymmetric data performed comparably when mapping from fine-grained to coarse-grained taxonomies, particularly in the case of Cognitive Tutor (using Skill2vec) to ASSISTments (using TAMF), which scored 0.88 in recall@5. 

These results are promising, but do not yet reach the level required for completely unattended automatic taxonomy mapping. The approach can, however, effectively triage a skill mapping or taxonomy crosswalk process and likely reduce the manual labor needed by a considerable amount. Use of the methods introduced in this paper, amplified by improvements defined in future work, could facilitate a more connected digital learning ecosystem, providing greater acknowledgement of students' prior learning and subsequently, more effective pedagogy.    

\bibliographystyle{ACM-Reference-Format}
\bibliography{sample-base}

\end{document}